\documentclass[a4paper,11pt]{article}
\usepackage[dvips]{graphicx}
\usepackage{amssymb}
\usepackage{amsmath}

\usepackage{cite}

\begin{document}

\title{Fermions in a Reissner-Nordstr\"{o}m-anti-de Sitter Black Hole Background and $N=4$ Supersymmetry with non-trivial Topological Charges}
\author{
V.K. Oikonomou\thanks{voiko@physics.auth.gr}\\
Department of Theoretical Physics, Aristotle University of Thessaloniki,\\
54124 Thessaloniki, Greece
} \maketitle

\begin{abstract}
We demonstrate that the fermions in Reissner-Nordstr\"{o}m-anti-de Sitter black hole background in the chiral limit $m=0$, are related to an $N=4$ extended one dimensional supersymmetry with non-trivial topological charges. We also show that the $N=4$ extended supersymmetry is unbroken and we extend the two $N=2$ one dimensional supersymmetries that the system also possesses, to have a trivial central charge. The implications of the trivial central charge on the Witten index are also discussed.
\end{abstract}

\section*{Introduction and Motivation}

Supersymmetric quantum mechanics \cite{reviewsusyqm0,reviewsusyqm1,reviewsusyqm2,reviewsusyqm3,reviewsusyqm4,reviewsusyqm5,reviewsusyqm6,reviewsusyqm7,thaller} was originally introduced in order to have a simplified model of supersymmetry breaking in four dimensional quantum field theory. Thereafter, supersymmetric quantum mechanics (which we abbreviate to SUSY QM hereafter) has developed to be an independent research field, with many interesting applications in various research areas (for an important stream of reviews and articles consult \cite{reviewsusyqm0,reviewsusyqm1,reviewsusyqm2,reviewsusyqm3,reviewsusyqm4,reviewsusyqm5,reviewsusyqm6,reviewsusyqm7}). Hilbert spaces properties corresponding to SUSY QM systems and also applications to various quantum mechanical systems were presented in \cite{diffgeomsusyduyalities} and \cite{various,susyqminquantumsystems1,susyqminquantumsystems2,plu11,plu12,plu13,plu14,plu15,plu16,plu17,plu18,plu19,plu120,plu21,plu22,plu23,plu24,plu25} respectively. Some applications of SUSY QM to scattering phenomena appear in \cite{susyqmscatter} and particular features of supersymmetry breaking were studied in \cite{susybreaking}. Furthermore, higher $N$-extended one dimensional supersymmetries and their relation to harmonic superspaces or gravity, were presented in \cite{extendedsusy1,extendedsusy2,ivanov1,ivanov2}. SUSY QM was firstly introduced to model supersymmetry breaking in quantum field theory in four dimensions \cite{witten1}. Supersymmetry is an important theoretical tool in quantum field theory and string theory, with various implications, applications and predictions both phenomenological and theoretical \cite{odi11,odi12,odi13,odi14,odi15,odi16,odi17,odi21,odi22,odi31,odi32,haag,wittentplc,fayet,topologicalcharges,topologicalcharges1,topologicalcharges2}. Since up to date there is no experimental verification of supersymmetry, it must be broken in our four dimensional world (for some elegant ways to introduce supersymmetry breaking consult \cite{odi11,odi12,odi13,odi14,odi15,odi16,odi17,odi21,odi22}). However, due to the fact that the representations of higher N-extended supersymmetry in four dimensions are difficult to find, simplified dimensionally reduced SUSY QM models can provide us with important insights on these problems.

In a previous paper \cite{oikonomoureissner} we demonstrated that the fermionic system in a Reissner-Nordstr\"{o}m-anti-de Sitter black hole gravitational background is related to two different $N=2$, $d=1$ supersymmetric quantum mechanical algebras with no central charges. In that paper we posed the question whether these two supersymmetries, which at first sight seem totally different, can be combined to form a more involved supersymmetric structure. In this paper we shall demonstrate that the answer lies in the affirmative and particularly that the enhanced supersymmetric structure is that of an $N=4$, $d=1$ supersymmetric algebra, with non-trivial topological charges. With topological charge we mean an operator equal to the anticommutator of Fermi charges (supercharges), following closely the terminology of references \cite{wittentplc,haag,fayet}. As we shall see, this $N=4$, $d=1$ supersymmetry is unbroken and we exclude the possibility of even partial supersymmetry breaking. In addition, in order to cover all possible non-trivial extensions of the existing supersymmetric structures, we present a central charge extended version of each $N=2$, $d=1$ supersymmetry and study the implications this central charge has on the Witten index of the $N=2$ system.

This paper is organized as follows: In section 1 we briefly present all the necessary theoretical framework concerning fermions in a Reissner-Nordstr\"{o}m-anti-de Sitter black hole gravitational background and their connection to $N=2$, $d=1$ supersymmetry, following reference \cite{oikonomoureissner}. In section 2, we present in detail the enhanced underlying $N=4$, $d=1$ supersymmetric structure of the system and in addition we study a kind of duality that the algebra possesses. A small discussion on non-trivial topological charges follows in the end of section 2. In section 3, we present a central charge extension of each $N=2$, $d=1$ supersymmetry and the implications of the trivial central charge on the Witten index. The conclusions follow in the end of the paper.

\section{$N=2$ SUSY QM and Massless Dirac Fermion Quasinormal Modes in Reissner-Nordstr\"{o}m-anti-de Sitter black hole spacetimes}

In this section we shall briefly review how $N=2$, $d=1$ supersymmetry is related to a system of Dirac fermions in a
Reissner-Nordstr\"{o}m-anti-de Sitter black hole gravitational background. For a detailed account on this issue consult reference \cite{oikonomoureissner}. The supersymmetry we shall present
shortly, is very closely related to the quasinormal modes \cite{roman} of the
Dirac fermionic field in the Reissner-Nordstr\"{o}m-anti-de Sitter
black hole background \cite{oikonomoureissner}.

\subsection{A Brief Review of $N=2$ SUSY QM}

In order to maintain the article self contained we shall briefly review the essentials of unbroken $N=2$ supersymmetric quantum mechanics algebra. The generators of the $N=2$ algebra are the two supercharges which we denote in general $Q_1$ and
$Q_2$ and also a Hamiltonian $H$. These operators obey \cite{reviewsusyqm0,reviewsusyqm1,reviewsusyqm2,reviewsusyqm3,reviewsusyqm4,reviewsusyqm5,reviewsusyqm6,reviewsusyqm7,thaller},
\begin{equation}\label{sxer2}
\{Q_1,Q_2\}=0,{\,}{\,}{\,}H=2Q_1^2=2Q_2^2=Q_1^2+Q_2^2
\end{equation}
The supercharges $Q_1$ and $Q_2$ can be used to define a new supercharge,
\begin{equation}\label{s2}
{\mathcal{Q}}=\frac{1}{\sqrt{2}}(Q_{1}+iQ_{2})
\end{equation}
and its adjoint,
\begin{equation}\label{s255}
{\mathcal{Q}}^{\dag}=\frac{1}{\sqrt{2}}(Q_{1}-iQ_{2})
\end{equation}
The new supercharges $\mathcal{Q}$ and $\mathcal{Q}$ satisfy,
\begin{equation}\label{s23}
Q^{2}={Q^{\dag}}^2=0
\end{equation}
and moreover,
\begin{equation}\label{s4}
\{{\mathcal{Q}},{\mathcal{Q}}^{\dag}\}=H
\end{equation}
A very important element of the supersymmetric algebra is the so-called Witten parity, $W$,
which is defined as,
\begin{equation}\label{s45}
[W,H]=0
\end{equation}
The Witten parity anti-commutes with the supercharges,
\begin{equation}\label{s5}
\{W,{\mathcal{Q}}\}=\{W,{\mathcal{Q}}^{\dag}\}=0
\end{equation}
and in addition it satisfies the following relation,
\begin{equation}\label{s6}
W^{2}=1
\end{equation}
The main utility of the Witten parity $W$, is that it actually spans the
Hilbert space $\mathcal{H}$ of the quantum system described by equations (\ref{s23}) and (\ref{s4}) to positive and
negative Witten parity subspaces, that is,
\begin{equation}\label{shoes}
\mathcal{H}^{\pm}=P_{s}^{\pm}\mathcal{H}=\{|\psi\rangle :
W|\psi\rangle=\pm |\psi\rangle \}
\end{equation}
Hence, the quantum system Hilbert space $\mathcal{H}$ can be
written $\mathcal{H}=\mathcal{H}^+\oplus \mathcal{H}^-$. For the
present case we shall choose a specific representation for the
operators defined above, which for the general case can be
represented as:
\begin{equation}\label{s7345}
W=\bigg{(}\begin{array}{ccc}
  I & 0 \\
  0 & -I  \\
\end{array}\bigg{)}
\end{equation}
with $I$ the $N\times N$ identity matrix. Bearing in mind that ${\mathcal{Q}}^2=0$
and $\{{\mathcal{Q}},W\}=0$, the supercharges can take the following form,
\begin{equation}\label{s7}
{\mathcal{Q}}=\bigg{(}\begin{array}{ccc}
  0 & A \\
  0 & 0  \\
\end{array}\bigg{)}
\end{equation}
and
\begin{equation}\label{s8}
{\mathcal{Q}}^{\dag}=\bigg{(}\begin{array}{ccc}
  0 & 0 \\
  A^{\dag} & 0  \\
\end{array}\bigg{)}
\end{equation}
Consequently the initial real supercharges $Q_1$ and $Q_2$ can be written,
\begin{equation}\label{s89}
Q_1=\frac{1}{\sqrt{2}}\bigg{(}\begin{array}{ccc}
  0 & A \\
  A^{\dag} & 0  \\
\end{array}\bigg{)}
\end{equation}
and also,
\begin{equation}\label{s10}
Q_2=\frac{i}{\sqrt{2}}\bigg{(}\begin{array}{ccc}
  0 & -A \\
  A^{\dag} & 0  \\
\end{array}\bigg{)}
\end{equation}
The $N\times N$ matrices $A$ and $A^{\dag}$, are in the general case operators which factorize the  $H_+$ and $H_-$ Hamiltonians and therefore these are the intertwining operators which generate the Darboux transformations between $H_+$ and $H_-$. In the case of a super-oscillator system, these operators are the bosonic creation-annihilation operators. These operators act as $A: \mathcal{H}^-\rightarrow
\mathcal{H}^+$ and $A^{\dag}:
\mathcal{H}^+\rightarrow \mathcal{H}^-$. Using relations
(\ref{s7345}), (\ref{s7}), (\ref{s8}) we can put the Hamiltonian $H$ into a diagonal form,
\begin{equation}\label{s11}
H=\bigg{(}\begin{array}{ccc}
  AA^{\dag} & 0 \\
  0 & A^{\dag}A  \\
\end{array}\bigg{)}
\end{equation}
Therefore, the total supersymmetric Hamiltonian $H$ that describes the
supersymmetric system, can be written in terms of the superpartner
Hamiltonians,
\begin{equation}\label{h1}
H_{+}=A{\,}A^{\dag},{\,}{\,}{\,}{\,}{\,}{\,}{\,}H_{-}=A^{\dag}{\,}A
\end{equation}
For reasons that will be immediately clear in the following, we define the operator
$P_s^{\pm}$, the eigenstates of which, $|\psi^{\pm}\rangle$, satisfy
the following relation:
\begin{equation}\label{fd1}
P^{\pm}_s|\psi^{\pm}\rangle =\pm |\psi^{\pm}\rangle
\end{equation}
Hence, we call them positive and negative parity eigenstates,
parity referring to the $P^{\pm}_s$ operator. Representing the
Witten operator as in (\ref{s7345}), the parity eigenstates can be
cast in the following representation,
\begin{equation}\label{phi5}
|\psi^{+}\rangle =\left(%
\begin{array}{c}
  |\phi^{+}\rangle \\
  0 \\
\end{array}%
\right)
\end{equation}
and also,
\begin{equation}\label{phi6}
|\psi^{-}\rangle =\left(%
\begin{array}{c}
  0 \\
  |\phi^{-}\rangle \\
\end{array}%
\right)
\end{equation}
with $|\phi^{\pm}\rangle$ $\epsilon$ $H^{\pm}$.

\subsection{Two $N=2$ SUSY QM Algebras for Fermions in Reissner-Nordstr\"{o}m-anti-de Sitter Spacetimes}

Let us now turn our focus to the case we have fermions in a Reissner-Nordstr\"{o}m-anti-de Sitter background. The metric in
a $d$-dimensional Reissner-Nordstr\"{o}m-anti-de Sitter spacetime is
given by:
\begin{equation}\label{RNmetric}
\mathrm{d}s^2=-f(r)\mathrm{d}t^2+\frac{1}{f(r)}\mathrm{d}r^2+r^2\mathrm{d}\Omega^2_{d-2}
\end{equation}
where the function $f(r)$ is equal to:
\begin{equation}\label{fr}
f(r)=1-\frac{2M}{r^{d-1}}+\frac{Q^2}{r^{2d-2}}+\frac{\Lambda}{3}r^2
\end{equation}
In the above equation, Q is the black hole
charge, $M$ is the black hole mass and $\Lambda$ the positive cosmological constant which characterizes the anti-de-Sitter spacetime. In addition, the cosmological constant is related to the anti-de-Sitter radius $R_0$, as $\Lambda =3/R_0^2$. Moreover, the
$\mathrm{d}\Omega^2_{d-2,k}$ is the metric of the $d-2$ dimensional sphere. In this paper we shall consider the case $d=4$. 
The spin connection in four dimensions $\omega_{\hat{a}\hat{b}c}$, is
equal to:
\begin{equation}\label{sup1}
\omega_{\hat{a}\hat{b}c}=e_{\hat{a}d}\partial_{c}e^{d}_{\hat{b}}+e_{\hat{a}d}e^{f}_{\hat{b}}\Gamma^{d}_{{\,}{\,}fc}
\end{equation}
where, $e_{\hat{a}d}$ denotes the tetrad field, while
$\Gamma^{d}_{{\,}{\,}fc}$ denotes the Christoffel connection. The
Einstein-Maxwell action for the Dirac fermion field equals to:
\begin{align}\label{actionrn}
&
\mathcal{S}=\frac{1}{2G_4^2}\int\mathrm{d}^2x\sqrt{-g}\Big{(}\mathcal{R}+2\Lambda \Big{)}
\\ \notag & +\mathcal{N}\int\mathrm{d}^4x\sqrt{-g}\Big{(}-\frac{1}{4}F_{ab}F^{ab}+i(\bar{\Psi}\Gamma^{\alpha}(D_a-iqA_a)\Psi-m\bar{\Psi}\Psi)\Big{)}
\end{align}
In the above action (\ref{actionrn}), $G_4$ is the 4-dimensional
gravitational constant, $\mathcal{R}$ is the corresponding Ricci
scalar, $\mathcal{N}$ is a total coefficient characterizing matter
fields, and $q$ is the coupling constant between the fermion field
and the abelian gauge field $A_a$. Notice that the cosmological constant $\Lambda >0$ appears with a positive sign, since it describes anti-de-Sitter spacetime. In addition, the operator
$D_a$ is:
\begin{equation}\label{sup2}
D_a=\partial_a+\frac{1}{2}\omega_{\hat{c}\hat{b}a}\Sigma^{\hat{c}\hat{b}}
\end{equation}
with
$\Sigma^{\hat{c}\hat{b}}=\frac{1}{4}[\Gamma^{\hat{c}},\Gamma^{\hat{b}}]$. Moreover the Dirac gamma matrices are related to the vierbeins as follows,
\begin{equation}
\Gamma^b=e^b_{\hat{a}}\Gamma^{\hat{a}}
\end{equation}
In order to consistently extract the quasinormal mode spectrum corresponding to
the Reissner-Nordstr\"{o}m-anti-de Sitter black hole spacetime, we
consider the limit in which the fermionic field does not backreact
on the metric. The wave function solution
$\Psi(r,x_{\mu},t)$ can be written in the following form:
\begin{equation}\label{functionpsi}
\Psi(r,x_{\mu},t)=\psi(r)e^{-i\omega t+i\vec{k}{\,}{\,}\cdot
\vec{x}}
\end{equation}
with $x_{\mu}=(x,y)$, $r=\sqrt{x^2+y^2}$, $\vec{x}=(x,y)$ and $\vec{k}=(k_x,k_y)$. Using the above form
of the wave function (\ref{functionpsi}), the Dirac equation can be
cast into the following form:
\begin{equation}\label{dequatafter}
\sqrt{f}\Gamma^{\hat{r}}\partial_r \psi-\frac{i
\omega}{\sqrt{f}}\Gamma^{\hat{t}}\psi+\frac{i\vec{k}\cdot
\Gamma^{\hat{\vec{x}}}}{r}\psi+\frac{1}{4}\Big{(}\frac{f'}{\sqrt{f}}+\frac{4\sqrt{f}}{r}\Big{)}\Gamma^{\hat{r}}\psi-(iq\Gamma^{\alpha}A_{\alpha}+m)\psi=0
\end{equation}
with $\vec{k}\cdot
\Gamma^{\hat{\vec{x}}}=k_x\Gamma^{\hat{x}}+k_y\Gamma^{\hat{\vec{y}}}$.
The Dirac gamma matrices can be written in the following
representation:
\begin{equation}\label{dgammama}
\Gamma^{\hat{t}}=\bigg{(}\begin{array}{ccc}
  I & 0 \\
  0 & -I  \\
\end{array}\bigg{)},{\,}{\,}{\,}{\,}\Gamma^{\hat{i}}=\bigg{(}\begin{array}{ccc}
  0 & \sigma^{\hat{i}} \\
  \sigma^{\hat{i}} & 0  \\
\end{array}\bigg{)}
\end{equation}
with $I$ the identity matrix and $\sigma^{i}$ the Pauli matrices,
namely:
\begin{equation}\label{sigmapauli}
\sigma^{\hat{x}}=\bigg{(}\begin{array}{ccc}
  0 & 1 \\
  1 & 0  \\
\end{array}\bigg{)},{\,}{\,}{\,}{\,}\sigma^{\hat{y}}=\bigg{(}\begin{array}{ccc}
  0 & -i \\
  i & 0  \\
\end{array}\bigg{)},{\,}{\,}\sigma^{\hat{r}}=\bigg{(}\begin{array}{ccc}
  1 & 0 \\
  0 & -1 \\
\end{array}\bigg{)}
\end{equation}
In order to simplify the problem at hand, we decompose the fermion field Hilbert
space to the chirality operator subspaces, that is:
\begin{equation}\label{chiral1}
\Psi=\bigg{(}\begin{array}{ccc}
  \Psi_{+} \\
  \Psi_{-}  \\
\end{array}\bigg{)}
\end{equation}
and $\mathcal{P}_{\pm}\Psi=\pm\Psi_{\pm}$, with $\mathcal{P}_{\pm}=\frac{1}{2}(1\pm \Gamma^5)$ and also
$\Gamma^5=i\Gamma^t\Gamma^x\Gamma^y\Gamma^r$. Using the eigenstates
$\Psi_{\pm}$, the Dirac equations of motion can be cast as:
\begin{equation}\label{deqm2}
(\sqrt{f}\partial_r+\frac{1}{4}\frac{f'}{\sqrt{f}}+\frac{\sqrt{f}}{r})\sigma^{\hat{r}}\psi_{-}\frac{i}{r}(\vec{k}\cdot
\vec{\sigma})\psi_{-}+\frac{i}{\sqrt{f}}(\omega+qA_t)\psi_{-}-m\psi_+=0
\end{equation}
 and in addition,
\begin{equation}\label{deqm22}
(\sqrt{f}\partial_r+\frac{1}{4}\frac{f'}{\sqrt{f}}+\frac{\sqrt{f}}{r})\sigma^{\hat{r}}\psi_{+}\frac{i}{r}(\vec{k}\cdot
\vec{\sigma})\psi_{+}+\frac{i}{\sqrt{f}}(\omega+qA_t)\psi_{-}-m\psi_{-}=0
\end{equation}
with $\Psi_+=\psi_+e^{-i\omega t+i\vec{k}\vec{x}}$ and
$\Psi_-=\psi_{-}e^{-i\omega t+i\vec{k}\vec{x}}$. Notice that the set of the
above equations (\ref{deqm2}) and (\ref{deqm22}) is invariant
under the transformation:
\begin{equation}\label{transfo}
\omega \rightarrow -\omega,{\,}{\,}{\,}{\,}{\,}q\rightarrow -q,
{\,}{\,}{\,}{\,}{\,}\psi_{+}\rightarrow \psi_{-}
\end{equation}
Side remark, these transformation properties of the equations (\ref{transfo}) are actually not inherited to the operators of the higher extended supersymmetries we shall present in the following sections, as we shall see in section 2. Remarkably however, the extended supersymmetric algebras are invariant under the transformation (\ref{transfo}).

In the rest of this paper we shall be interested in the chiral
limit of the fermionic system, which means that the fermion mass is zero, that is $m=0$. This will result to an unbroken chiral symmetry for
the system. We also focus on the
quasinormal modes of $\psi_{+}$, since the analysis for the $\psi_{-}$ is similar and we shall present only the results for $\psi_{-}$. Moreover we set $k_y=0$ because of
the symmetry that the system possesses on the
$(\vec{x},\vec{y})$-plane. Rewriting $\psi_{+}$ as
$\psi_{+}=r^{-1}f^{-1/4}\tilde{\psi}$, the equation (\ref{deqm22})
can be cast to the following form:
\begin{equation}\label{simeqd}
\sigma^{\hat{r}}\tilde{\psi}'-\frac{i}{f}(\omega+qA_t-\frac{\sqrt{f}}{r}k_{x}\sigma^{\hat{x}})\tilde{\psi}=0
\end{equation}
Further decomposing the fermionic field $\tilde{\psi}$, as:
\begin{equation}\label{decomp}
\tilde{\psi}=\bigg{(}\begin{array}{ccc}
  \psi_{1} \\
  \psi_{2}  \\
\end{array}\bigg{)}
\end{equation}
the equation (\ref{simeqd}), can be recast in the following
form:
\begin{align}\label{decver222}
&
\psi_1'-\frac{i}{f}(\omega+qA_t)\psi_1+\frac{i}{r\sqrt{f}}k_x\psi_2=0
\\ \notag &
\psi_2'+\frac{i}{f}(\omega+qA_t)\psi_2-\frac{i}{r\sqrt{f}}k_{x}\psi_1=0
\end{align}
The above equations are invariant under the symmetry:
\begin{equation}\label{transfo2}
\omega \rightarrow -\omega,{\,}{\,}{\,}{\,}{\,}q\rightarrow -q,
{\,}{\,}{\,}{\,}{\,}k_{x}\rightarrow -k_{x},
{\,}{\,}{\,}{\,}{\,}\psi_{1}\rightarrow \psi_{2}
\end{equation}
As we already mentioned earlier, this symmetry is the origin of the symmetry that the higher $N=4$ extended SUSY QM algebra possesses, as we shall see in section 2. Using the equations (\ref{decver222}) we can construct an
$N=2$ supersymmetric algebra. This algebra is founded on the
matrix:
\begin{equation}\label{susyqmrn567m12}
{{{{{\mathcal{D}}_{RN}}}}}=\left(%
\begin{array}{cc}
 \partial_r-\frac{i}{f}(\omega+qA_t) & \frac{i}{r\sqrt{f}}k_x
 \\  -\frac{i}{r\sqrt{f}}k_x & \partial_r+\frac{i}{f}(\omega+qA_t)\\
\end{array}%
\right)
\end{equation}
acting on the vector:
\begin{equation}\label{ait34e1}
\left(%
\begin{array}{c}
 \psi_1 \\
  \psi_2 \\
\end{array}%
\right)
\end{equation}
As can easily be seen, the zero modes of the matrix
(\ref{susyqmrn567m12}) yield the solutions of equation
(\ref{decver222}) with respect to the parameter $\omega$ which solutions
correspond to the zero modes of the Dirac fermionic system.
Thereby, the zero mode solutions of the matrix
(\ref{susyqmrn567m12}) and the quasinormal modes of the Dirac
fermionic system are in bijective correspondence. Hence, the
existence of quasinormal modes guarantees the existence of zero
modes for the aforementioned matrix $\mathcal{D}_{RN}$. The adjoint of the matrix
${{{{{\mathcal{D}}_{RN}}}}}$ is equal to:
\begin{equation}\label{adjsusyqmrn567m}
{{{{{\mathcal{D}}_{RN}}}}}^{\dag}=\left(%
\begin{array}{cc}
 -\partial_r+\frac{i}{f}(\omega^*+qA_t) & \frac{i}{r\sqrt{f}}k_x
 \\  -\frac{i}{r\sqrt{f}}k_x & -\partial_r-\frac{i}{f}(\omega^*+qA_t)\\
\end{array}%
\right)
\end{equation}
Correspondingly, the supercharges of the $N=2$ algebra ${\mathcal{Q}}_{RN}$
and $Q^{\dag}_{RN}$  are equal to:
\begin{equation}\label{wit2dr}
{\mathcal{Q}}_{RN}=\bigg{(}\begin{array}{ccc}
  0 & {{{{{\mathcal{D}}_{RN}}}}} \\
  0 & 0  \\
\end{array}\bigg{)}, {\,}{\,}{\,}{\,}{\,}Q^{\dag}_{RN}=\bigg{(}\begin{array}{ccc}
  0 & 0 \\
  {{{{{\mathcal{D}}_{RN}}}}}^{\dag} & 0  \\
\end{array}\bigg{)}
\end{equation}
In addition, the quantum Hamiltonian is equal to,
\begin{equation}\label{wit4354dr}
H_{RN}=\bigg{(}\begin{array}{ccc}
  {{{{\mathcal{D}}_{RN}}}}{{{{\mathcal{D}}_{RN}}}}^{\dag} & 0 \\
  0 & {{{{\mathcal{D}}_{RN}}}}^{\dag}{{{{\mathcal{D}}_{RN}}}}  \\
\end{array}\bigg{)}
\end{equation}
The supercharges (\ref{wit2dr}) and the Hamiltonian and
(\ref{wit4354dr}), satisfy the equations (\ref{s23}) and
(\ref{s45}), namely,
\begin{equation}\label{structureqns}
\{{\mathcal{Q}}_{RN},{\mathcal{Q}}_{RN}^{\dag}\}=H_{RN},{\,}{\,}{\mathcal{Q}}_{RN}^2=0,{\,}{\,}{{\mathcal{Q}}_{RN}^{\dag}}^2=0,{\,}{\,}\{{\mathcal{Q}}_{RN},W\}=0,{\,}{\,}W^2=I,{\,}{\,}[W,H_{RN}]=0
\end{equation}
Therefore the algebraic structure of an $N=2$ SUSY QM algebra underlies this fermionic system that corresponds to the solution $\psi_+$ (recall that there is another identical system corresponding to $\psi_{-}$, which we shall describe soon). Let us see if this underlying supersymmetry is broken or unbroken. One easy way to see whether supersymmetry is broken or not is to calculate the index of the operator ${{{{\mathcal{D}}_{RN}}}}$.

Since the number of quasinormal modes forms a discrete infinite set, and due to the bijective correspondence between the zero modes of the operator ${\mathcal{D}}_{RN}$ and the quasinormal modes, we conclude that the zero modes form a discrete infinite set. Hence, the operator ${\mathcal{D}}_{RN}$ is not Fredholm which means that the index of this operator and correspondingly the Witten index must be regularized.
In order to do so, we shall make use of the heat-kernel
regularized index \cite{reviewsusyqm0,reviewsusyqm1,reviewsusyqm2,reviewsusyqm3,reviewsusyqm4,reviewsusyqm5,reviewsusyqm6,reviewsusyqm7,thaller}, for the operator ${\mathcal{D}}_{RN}$, denoted
$\mathrm{ind}_t{\mathcal{D}}_{RN}$ and also for the Witten index, $\Delta_t$. These regularized indices are defined as follows:
\begin{align}\label{heatkerw}
&
\mathrm{ind}_t{\mathcal{D}}_{RN}=\mathrm{Tr}(-We^{-t{\mathcal{D}}_{RN}^{\dag}{\mathcal{D}}_{RN}})=\mathrm{tr}_{-}(-We^{-t{\mathcal{D}}_{RN}^{\dag}{\mathcal{D}}_{RN}})-\mathrm{tr}_{+}(-We^{-t{\mathcal{D}}_{RN}{\mathcal{D}}_{RN}^{\dag}})
\\ \notag & \Delta_t=\lim_{t\rightarrow
\infty}\mathrm{ind}_t{\mathcal{D}}_{RN}
\end{align}
The parameter $t$, is a positive number $t>0$, and in addition the
trace $\mathrm{tr}_{\pm }$, stands for the trace in the subspaces
$\mathcal{H}^{\pm}$ we defined in the previous subsection. The heat-kernel regularized index is defined only
for trace class operators \cite{thaller}. In the regularized
index case, the same hold true, in reference to supersymmetry breaking,
that is, if $\mathrm{\Delta}_t\neq 0$ then supersymmetry is unbroken. In the case
the Witten index is zero, if
$\mathrm{ker}{\mathcal{D}}_{RN}=\mathrm{ker}{\mathcal{D}}_{RN}^{\dag}=0$, then supersymmetry is
broken, while when
$\mathrm{ker}{\mathcal{D}}_{RN}=\mathrm{ker}{\mathcal{D}}_{RN}^{\dag}\neq0$ supersymmetry
is unbroken. In our case, supersymmetry is unbroken and we
can see this without solving the zero mode equation of the
${\mathcal{D}}_{RN}^{\dag}$ operator. Indeed, the existence of zero modes
suffices to argue about supersymmetry. Since
$\mathrm{ker}{\mathcal{D}}_{RN}\neq 0$,  the zero modes equation for the
operator ${\mathcal{D}}_{RN}^{\dag}$ can yield two results. Either that
$\mathrm{ker}{\mathcal{D}}_{RN}^{\dag}\neq 0$ or that
$\mathrm{ker}{\mathcal{D}}_{RN}^{\dag}=0$. If the latter is true, then the
Witten index is different than zero, $\Delta_t\neq 0$, which implies that
supersymmetry is unbroken. In the former case,
$\mathrm{ker}{\mathcal{D}}_{RN}^{\dag}\neq 0$, which can either mean that
$\mathrm{ker}{\mathcal{D}}_{RN}^{\dag}=\mathrm{ker}{\mathcal{D}}_{RN}$ or that
$\mathrm{ker}{\mathcal{D}}_{RN}^{\dag}\neq \mathrm{ker}{\mathcal{D}}_{RN}$. In both the aforementioned cases
supersymmetry is unbroken. Therefore, the system that is described by
the $\psi_{+}$ function possesses an unbroken underlying $N=2$
supersymmetry.

Recall that there is another solution to the Dirac equation in
this curved background, namely $\psi_{-}$ (see below equation (\ref{deqm22})). The equations of motion
corresponding to $\psi_{-}$ are equal to:
\begin{align}\label{decver2}
&
\psi_1'-\frac{i}{f}(-\omega-qA_t)\psi_1'-\frac{i}{r\sqrt{f}}k_x\psi'_2=0
\\ \notag &
\psi_2'+\frac{i}{f}(-\omega-qA_t)\psi_2'+\frac{i}{r\sqrt{f}}k_{x}\psi_1'=0
\end{align}
where we introduced the notation,
\begin{equation}\label{decomp1}
\tilde{\psi}'=\bigg{(}\begin{array}{ccc}
  \psi_{1}' \\
  \psi_{2}'  \\
\end{array}\bigg{)}
\end{equation}
and $\psi_{-}$ being related to
$\psi_{-}=r^{-1}f^{-1/4}\tilde{\psi}'$. By the same token as
in the $\psi_{+}$ case, the supersymmetric quantum algebra can be
built by using the matrix:
\begin{equation}\label{susyqmrn567m}
{\mathcal{D}}_{RN'}=\left(%
\begin{array}{cc}
 \partial_r-\frac{i}{f}(-\omega-qA_t) & -\frac{i}{r\sqrt{f}}k_x
 \\  \frac{i}{r\sqrt{f}}k_x & \partial_r+\frac{i}{f}(-\omega-qA_t)\\
\end{array}%
\right)
\end{equation}
acting on the vector:
\begin{equation}\label{ait34e1}
\left(%
\begin{array}{c}
 \psi_1' \\
  \psi_2' \\
\end{array}%
\right)
\end{equation}
The supercharges of the new algebra are equal to,
\begin{equation}\label{wit2dr1}
{\mathcal{Q}}_{RN'}=\bigg{(}\begin{array}{ccc}
  0 & {\mathcal{D}}_{RN'} \\
  0 & 0  \\
\end{array}\bigg{)}, {\,}{\,}{\,}{\,}{\,}Q^{\dag}_{RN'}=\bigg{(}\begin{array}{ccc}
  0 & 0 \\
  {\mathcal{D}}_{RN'}^{\dag} & 0  \\
\end{array}\bigg{)}
\end{equation}
and moreover the Hamiltonian is,
\begin{equation}\label{wit4354dr1231}
H_{RN'}=\bigg{(}\begin{array}{ccc}
  {\mathcal{D}}_{RN'}{\mathcal{D}}_{RN'}^{\dag} & 0 \\
  0 & {\mathcal{D}}_{RN'}^{\dag}{\mathcal{D}}_{RN'}  \\
\end{array}\bigg{)}
\end{equation}
For the same reasoning as in the previous, we find an $N=2$ underlying supersymmetry in this case too. Denoting the algebra
corresponding to $\psi_{-}$, $N_2$ and the one corresponding to
$\psi_1$, $N_1$, we have come to the result that the Dirac
fermionic system in an Reissner-Nordstr\"{o}m-anti-de Sitter
background, possesses an total supersymmetry $N$, which is the direct sum
of two $N=2$ supersymmetries, namely:
\begin{equation}\label{directsum}
N_{total}=N_1\oplus N_2
\end{equation}
Therefore the total Hamiltonian of the system is $H_{total}=H_{RN'}+H_{RN}$.
It is the subject of this paper to investigate if this supersymmetry $N_{total}$
results after the breaking of a larger supersymmetry, for example
an $N=4$ supersymmetry, or even the possibility that a central
charge exists.

As we shall evince in the next section, this $N_{total}$ supersymmetry is actually an indication that there exists a richer unbroken supersymmetric structure underlying the system. This underlying supersymmetry is  an $N=4$ supersymmetric quantum mechanics algebra with non trivial topological charges and not central charges. 
 
\subsection{Comment on SUSY QM and Global Spacetime Supersymmetry}

With the additional structure of a SUSY QM algebra on the fermionic degrees of freedom, the spacetime is rendered a supermanifold, at least locally but we refer the interested reader to reference \cite{oiko2} where this issue is studied in detail. With the spacetime manifold being locally a supermanifold, it is unavoidable to seek any possible connection of the SUSY QM algebra with a global spacetime supersymmetry. However, there exists no such connection as we now discuss. When someone studies supersymmetric algebras in various spacetime dimensions other than $d=1$, has to be cautious since the four dimensional spacetime supersymmetric algebra (which is a graded super-Poincare algebra in four dimensions) is four dimensional while the SUSY QM algebra is actually one dimensional. Moreover, spacetime global supersymmetry in $d>1$ dimensions and
SUSY QM, that is $d=1$ supersymmetry, are in principle different, although by dimensional reduction of certain four dimensional Super-Yang-Mills models one may obtain one dimensional supersymmetric sigma models. Particularly, extended (with $N = 4, 6...$) SUSY QM models are the result of the dimensional reduction to one dimension of $N = 2$ and $N = 1$
Super-Yang Mills models. However, no $N=2$ SUSY QM can be obtained this way. Actually, the $N = 2$, $d=1$ SUSY QM supercharges
do not generate spacetime supersymmetry and thereby, SUSY QM does not relate fermions and bosons in terms of representations of the super-Poincare algebra in four dimensions. Particularly, the SUSY QM supercharges provide the Hilbert space of quantum states of the system under study, with a $Z_2$ grading and in addition they generate transformations between the Witten parity eigenstates. This is actually the reason why the manifold on which fermions are defined is globally a graded manifold and not a supermanifold (see \cite{oiko2} for details).
 
\section{$N=4$ Extended $d=1$ Supersymmetry with non-trivial Topological Charges}

In this section we shall demonstrate that the two $N=2$ supersymmetries described in the previous section combine to form a $N=4$, $d=1$ supersymmetric algebra with non-trivial topological charges. As topological charges we shall consider operators that are equal to the anticommutator of some supercharges, exactly as topological charges are considered in references \cite{wittentplc,haag,fayet}. The topological charges which we shall present do not commute with all the operators of the algebra, a feature quite often met in string theory contexts with supercharges. For example when a Wess-Zumino term is taken into account in a Ad$S_5\times S^5$ D-brane background with superalgebra $su(2,2|4)$ \cite{topologicalcharges2}, with the latter being the super-isometry algebra of the Ad$S_5\times S^5$ background. 

The system of fermions in the Reissner-Nordstr\"{o}m-anti-de Sitter has the following two complex supercharges,  
\begin{equation}\label{wit2jdnhdsoft}
\mathcal{Q}_{ RN}=\bigg{(}\begin{array}{ccc}
  0 & D_{RN } \\
  0 & 0  \\
\end{array}\bigg{)},{\,}{\,}{\,}
\mathcal{Q}_{RN'}=\bigg{(}\begin{array}{ccc}
  0 & D_{RN'} \\
  0 & 0  \\
\end{array}\bigg{)}
\end{equation}
with the operators $\mathcal{D}_{RN}$ and $\mathcal{D}_{RN'}$ defined in relations (\ref{susyqmrn567m12}) and (\ref{susyqmrn567m}). For notational simplicity we introduce the following two operators $\mathcal{L}_1$ and $\mathcal{L}_2$
\begin{equation}\label{operatorsintroduced}
\mathcal{L}_1=\frac{1}{f}(\omega+qA_t),{\,}{\,}{\,}\mathcal{L}_2=\frac{1}{f}(\omega^*+qA_t)
\end{equation}
Notice that $\mathcal{L}_1^{\dag}=\mathcal{L}_2$ and also there is no physical sense in these operators, we just introduced these in order to simplify the equations in the rest of this section. In order to reveal the underlying $N=4$, $d=1$ supersymmetric structure of the fermionic quantum system, we compute the following commutation and anticommutation relations:
\begin{align}\label{commutatorsanticomm}
&\{{{\mathcal{Q}}_{RN'}},{{\mathcal{Q}}_{RN'}}^{\dag}\}=2\mathcal{H}+\mathcal{Z}_{RN'RN'},
{\,}\{{{\mathcal{Q}}_{RN}},{{\mathcal{Q}}_{RN}}^{\dag}\}=2\mathcal{H}+\mathcal{Z}_{RNRN}
,{\,}\{{{\mathcal{Q}}_{RN}},{{\mathcal{Q}}_{RN}}\}=0,
\\ \notag &\{{{\mathcal{Q}}_{RN'}},{{\mathcal{Q}}_{RN'}}\}=0, {\,} \{{{\mathcal{Q}}_{RN}},{{\mathcal{Q}}_{RN'}}^{\dag}\}=\mathcal{Z}_{RNRN'},{\,}\{{{\mathcal{Q}}_{RN'}},{{\mathcal{Q}}_{RN}}^{\dag}\}=\mathcal{Z}_{RN' RN},{\,}\\ \notag
&\{{{\mathcal{Q}}_{RN'}}^{\dag},{{\mathcal{Q}}_{RN'}}^{\dag}\}=0,\{{{\mathcal{Q}}_{RN}}^{\dag},{{\mathcal{Q}}_{RN}}^{\dag}\}=0,{\,}\{{{\mathcal{Q}}_{RN}}^{\dag},{{\mathcal{Q}}_{RN'}}^{\dag}\}=0,{\,}\{{{\mathcal{Q}}_{RN}},{{\mathcal{Q}}_{RN'}}\}=0{\,}\\
\notag
&[{{\mathcal{Q}}_{RN'}},{{\mathcal{Q}}_{RN}}]=0,[{{\mathcal{Q}}_{RN}}^{\dag},{{\mathcal{Q}}_{RN'}}^{\dag}]=0,{\,}[{{\mathcal{Q}}_{RN'}},{{\mathcal{Q}}_{RN'}}]=0,{\,}[{{\mathcal{Q}}_{RN'}}^{\dag},{{\mathcal{Q}}_{RN'}}^{\dag}]=0
\end{align}
Let us describe in detail the new operators that appear in the above relation (\ref{commutatorsanticomm}). The operator $\mathcal{H}$ stands for:
\begin{equation}\label{newsusymat1a}
\mathcal{H}=\Big{(}\partial_r^2+\frac{k_x^2}{r^2f}+\mathcal{L}_1\mathcal{L}_2\Big{)}\times \mathrm{diag}\left ( \begin{array}{ccccc}1,1,1,1\end{array}\right )
\end{equation}
In addition, the operators $\mathcal{Z}_{RN RN}$ and $\mathcal{Z}_{RN' RN'}$ are equal to:
\begin{equation}\label{newsusymat2a}
\mathcal{Z}_{RN RN}=\left ( \begin{array}{ccccc}
  \mathcal{Z}^1_{RN RN} & 0 \\
  0 & \mathcal{Z}^2_{RN RN}  \\
\end{array}\right ),{\,}{\,}{\,}\mathcal{Z}_{ RN' RN' }=\left ( \begin{array}{ccccc}
  \mathcal{Z}^1_{RN' RN' } & 0 \\
  0 & \mathcal{Z}^2_{RN' RN' }  \\
\end{array}\right )
\end{equation}
with the operator $\mathcal{Z}^1_{RN RN}$ being equal to,
\begin{equation}\label{newsusymat3a}
\mathcal{Z}^1_{RN RN}=\left ( \begin{array}{cc}
 i\partial_r\mathcal{L}_2-i\mathcal{L}_1\partial_r & i(\partial_r-i\mathcal{L}_1)\frac{k_x}{r\sqrt{f}}+\frac{ik_x}{r\sqrt{f}}(\partial_r-i\mathcal{L}_2) \\
 -i(\partial_r+i\mathcal{L}_2)\frac{k_x}{r\sqrt{f}}-i(\partial_r+i\mathcal{L}_1)\frac{k_x}{r\sqrt{f}} &  -i\partial_r\mathcal{L}_2+i\mathcal{L}_1\partial_r \\
\end{array}\right )
\end{equation}
and additionally the operator $\mathcal{Z}^2_{RN RN}$ is equal to,
\begin{equation}\label{newsusymat4a}
\mathcal{Z}^2_{RN RN}=\left ( \begin{array}{cc}
  -i\partial_r\mathcal{L}_1+i\mathcal{L}_2\partial_r & i(\partial_r+i\mathcal{L}_2)\frac{k_x}{r\sqrt{f}}+\frac{ik_x}{r\sqrt{f}}(\partial_r+i\mathcal{L}_1) \\
 -i(\partial_r-i\mathcal{L}_1)\frac{k_x}{r\sqrt{f}}-i(\partial_r-i\mathcal{L}_2)\frac{k_x}{r\sqrt{f}} &  i\partial_r\mathcal{L}_1-i\mathcal{L}_2\partial_r \\
\end{array}\right )
\end{equation}
Moreover, the operator $\mathcal{Z}^1_{RN' RN'}$ is equal to: 
\begin{equation}\label{newsusymat3asoldiers}
\mathcal{Z}^1_{RN' RN'}=\left ( \begin{array}{cc}
 -i\partial_r\mathcal{L}_2+i\mathcal{L}_1\partial_r & -i(\partial_r+i\mathcal{L}_1)\frac{k_x}{r\sqrt{f}}-\frac{ik_x}{r\sqrt{f}}(\partial_r+i\mathcal{L}_2) \\
 i\frac{k_x}{r\sqrt{f}}(\partial_r-i\mathcal{L}_2)+i(\partial_r-i\mathcal{L}_2)\frac{k_x}{r\sqrt{f}} &  i\partial_r\mathcal{L}_2-i\mathcal{L}_1\partial_r \\
\end{array}\right )
\end{equation}
and also $\mathcal{Z}^2_{RN' RN'}$ stands for:
\begin{equation}\label{newsusymat4asoldiers}
\mathcal{Z}^2_{RN' RN'}=\left ( \begin{array}{cc}
  i\partial_r\mathcal{L}_1-i\mathcal{L}_2\partial_r & -i(\partial_r-i\mathcal{L}_2)\frac{k_x}{r\sqrt{f}}-\frac{ik_x}{r\sqrt{f}}(\partial_r-i\mathcal{L}_1) \\
 i\frac{k_x}{r\sqrt{f}}(\partial_r+i\mathcal{L}_1)+i(\partial_r+i\mathcal{L}_2)\frac{k_x}{r\sqrt{f}} &  -i\partial_r\mathcal{L}_1+i\mathcal{L}_2\partial_r \\
\end{array}\right )
\end{equation}
Finally, the operator $\mathcal{Z}_{RN RN'}$ is equal to:
\begin{equation}\label{newsusymat5a}
\mathcal{Z}_{RN RN'}=\left ( \begin{array}{ccccc}
  \mathcal{Z}^1_{RN RN'} & 0 \\
  0 & \mathcal{Z}^2_{RN RN'  }  \\
\end{array}\right )
\end{equation}
with the operator $\mathcal{Z}^1_{RN RN'}$ standing for:
\begin{equation}\label{newsusymat6a}
\mathcal{Z}^1_{RN RN'}=\left ( \begin{array}{cc}
  (\partial_r-i\mathcal{L}_2)(\partial_r-i\mathcal{L}_1)-\frac{k_x^2}{r^2f} & -i(\partial_r-i\mathcal{L}_1)\frac{k_x}{r\sqrt{f}}+\frac{ik_x}{r\sqrt{f}}(\partial_r+i\mathcal{L}_2) \\
 -i\frac{k_x}{r\sqrt{f}}(\partial_r-i\mathcal{L}_2)+i(\partial_r+i\mathcal{L}_1)\frac{k_x}{r\sqrt{f}} &  (\partial_r+i\mathcal{L}_1)(\partial_r+i\mathcal{L}_2)-\frac{k_x^2}{r^2f} \\
\end{array}\right )
\end{equation}
and in addition the operator $\mathcal{Z}^2_{RN RN'}$ is equal to:
\begin{equation}\label{newsusymat7a}
\mathcal{Z}^2_{RN RN'}=\left ( \begin{array}{cc}
 (\partial_r-i\mathcal{L}_2)(\partial_r-i\mathcal{L}_1)-\frac{k_x^2}{r^2f} & i(\partial_r-i\mathcal{L}_2)\frac{k_x}{r\sqrt{f}}-\frac{ik_x}{r\sqrt{f}}(\partial_r+i\mathcal{L}_1) \\
 i\frac{k_x}{r\sqrt{f}}(\partial_r-i\mathcal{L}_1)-i(\partial_r+i\mathcal{L}_2)\frac{k_x}{r\sqrt{f}} &  (\partial_r+i\mathcal{L}_2)(\partial_r+i\mathcal{L}_1)-\frac{k_x^2}{r^2f} \\
\end{array}\right )
\end{equation}
The operator $\mathcal{Z}_{RN' RN}$ is easily found since $\mathcal{Z}_{RN' RN}=\mathcal{Z}_{RN RN'}^{\dag}$. The operators $\mathcal{Z}_{RN RN},\mathcal{Z}_{RN RN'}$ and $\mathcal{Z}_{RN' RN'}$, are equal to anticommutators of supercharges and consequently these are non-trivial topological charges \cite{wittentplc,fayet,topologicalcharges,topologicalcharges1} but not central charges in any case. The commutation and anticommutation relations appearing in (\ref{commutatorsanticomm}) describe an $N=4$, $d=1$ superalgebra with non-trivial topological charges $\mathcal{Z}_{RN RN},\mathcal{Z}_{RN RN'}$ an $\mathcal{Z}_{RN' RN'}$. Therefore, apart from the two different $N=2$ SUSY QM algebras underlying the fermionic system in the Reissner-Nordstr\"{o}m-anti-de Sitter curved background, there exists a richer supersymmetric algebra with non-trivial topological charges.

Noticeably, the topological charges that appear in equations (\ref{newsusymat2a}) and (\ref{newsusymat5a}) and also the Hamiltonian of the $N=4$ algebra (\ref{newsusymat1a}) are Hermitian operators. On the contrary the supercharges (\ref{wit2jdnhdsoft}) are not Hermitian. This result verifies the validity of the $N=4$ superalgebra (\ref{commutatorsanticomm}), since as can be easily checked the anticommutator of two supercharges is Hermitian. For example, the anticommutator $\{{{\mathcal{Q}}_{RN}},{{\mathcal{Q}}_{RN}}^{\dag}\}$ is by definition equal to:
\begin{equation}\label{wit2jdnhdsoftexample}
\{{{\mathcal{Q}}_{RN}},{{\mathcal{Q}}_{RN}}^{\dag}\}=\bigg{(}\begin{array}{ccc}
  D_{RN }D_{RN }^{\dag} & 0 \\
  0 & D_{RN }^{\dag}D_{RN }  \\
\end{array}\bigg{)}
\end{equation}
which is Hermitian. As we evinced earlier the following relation holds true:
\begin{equation}
 \{{{\mathcal{Q}}_{RN}},{{\mathcal{Q}}_{RN}}^{\dag}\}=2\mathcal{H}+\mathcal{Z}_{RNRN}
\end{equation}
Therefore, owing to the fact that the left hand side of equation (\ref{wit2jdnhdsoftexample}) is Hermitian, we expect the right hand side to Hermitian too, which is actually, since $\mathcal{H}$ and also $\mathcal{Z}_{RNRN}$ are Hermitian. The same applies for all the other commutators and anticommutators. This actually exemplifies that the algebra we revealed is a valid self-consistent underlying algebra of the system the physical implications of which should be further studied . We defer this task to a future work.

Before closing this section it worths discussing the issue of partial supersymmetry breaking in the context of $N=4$, $d=1$ supersymmetry. The $N=4$, $d=1$ supersymmetry is the simple case of one dimensional supersymmetry that makes sense to discuss about partial supersymmetry or spontaneous supersymmetry breaking \cite{ivanovclass}. Particularly, it would be possible to have partial supersymmetry breaking if the algebra contains a topological charge, for example $\mathcal{Z}_{RNRN}$ and the following relations would hold true \cite{ivanovclass},
\begin{align}\label{partialsupercharge}
&  \{{{\mathcal{Q}}_{RN}},{{\mathcal{Q}}_{RN}}^{\dag}\}=2\mathcal{H}+\mathcal{Z}_{RNRN} \\ \notag &
 \{{{\mathcal{Q}}_{RN'}},{{\mathcal{Q}}_{RN'}}^{\dag}\}=2\mathcal{H}-\mathcal{Z}_{RNRN}
\end{align}
with all the remaining commutators and anticommutators being zero. The operators ${{\mathcal{Q}}_{RN}},{{\mathcal{Q}}_{RN'}}$ are the complex supercharges that constitute the $N=4$ superalgebra and relation (\ref{partialsupercharge}) describes the partial supersymmetry breaking from one $N=4$ to two $N=2$, $d=1$ supersymmetries \cite{ivanovclass}. It is obvious that in our case the $N=4$, $d=1$ supersymmetry is intact for three reasons:
\begin{itemize}

\item First of all $\{{{\mathcal{Q}}_{RN'}},{{\mathcal{Q}}_{RN'}}^{\dag}\}=2\mathcal{H}+\mathcal{Z}_{RN'RN'}$ with $\mathcal{Z}_{RN'RN'} \neq \mathcal{Z}_{RNRN}$

\item There exist other non-zero anticommutation relations that produce non-trivial topological charges and particularly the following

\begin{equation}\label{rgido}
\{{{\mathcal{Q}}_{RN}},{{\mathcal{Q}}_{RN'}}^{\dag}\},{\,}{\,}{\,}\{{{\mathcal{Q}}_{RN'}},{{\mathcal{Q}}_{RN}}^{\dag}\}
\end{equation}

\end{itemize}
Therefore the $N=4$ supersymmetry of the fermionic system under study is unbroken.

\subsection{Invariance of the $N=4$ Supersymmetric under a Global non-Trivial Transformation}

The parameters $\omega$, $q$ describe the quasinormal mode frequency and the charge of the fermion as we saw in subsection 1.2, see also \cite{oikonomoureissner}. Consider the transformation of relation (\ref{transfo}), that is,
\begin{align}\label{transfduality}
&\omega\rightarrow -\omega \\ \notag & 
q \rightarrow -q \\ \notag &
k_x\rightarrow -k_x
\end{align}
As can be easily checked, the operator $\mathcal{D}_{RN}$ is not invariant under the transformation (\ref{transfduality}),  but transforms in the following way:
\begin{equation}\label{transd1}
\mathcal{D}_{RN}\rightarrow \mathcal{D}_{RN'}
\end{equation}
Therefore, it worths seeing how the $N=4$, $d=1$ supersymmetric algebra of relation (\ref{commutatorsanticomm}) transforms. By using the transformation (\ref{transd1}), the anticommutators are transformed in the following way: 
\begin{align}\label{transdual2}
&\{{{\mathcal{Q}}_{RN'}},{{\mathcal{Q}}_{RN'}}^{\dag}\} \rightarrow \{{{\mathcal{Q}}_{RN}},{{\mathcal{Q}}_{RN}}^{\dag}\},{\,}{\,}{\,}\{{{\mathcal{Q}}_{RN}},{{\mathcal{Q}}_{RN}}^{\dag}\} \rightarrow \{{{\mathcal{Q}}_{RN'}},{{\mathcal{Q}}_{RN'}}^{\dag}\} \\ \notag &
\{{{\mathcal{Q}}_{RN}},{{\mathcal{Q}}_{RN}}\} \rightarrow \{{{\mathcal{Q}}_{RN'}},{{\mathcal{Q}}_{RN'}}\},{\,}{\,}{\,}\{{{\mathcal{Q}}_{RN'}},{{\mathcal{Q}}_{RN'}}\} \rightarrow \{{{\mathcal{Q}}_{RN}},{{\mathcal{Q}}_{RN}}\} \\ \notag &
 \{{{\mathcal{Q}}_{RN}},{{\mathcal{Q}}_{RN'}}^{\dag}\}\rightarrow \{{{\mathcal{Q}}_{RN'}},{{\mathcal{Q}}_{RN}}^{\dag}\}{\,}{\,}{\,} \{{{\mathcal{Q}}_{RN'}},{{\mathcal{Q}}_{RN}}^{\dag}\}\rightarrow \{{{\mathcal{Q}}_{RN}},{{\mathcal{Q}}_{RN'}}^{\dag}\} \\ \notag &
\{{{\mathcal{Q}}_{RN'}}^{\dag},{{\mathcal{Q}}_{RN'}}^{\dag}\} \rightarrow \{{{\mathcal{Q}}_{RN}}^{\dag},{{\mathcal{Q}}_{RN}}^{\dag}\}{\,}{\,}{\,}\{{{\mathcal{Q}}_{RN}}^{\dag},{{\mathcal{Q}}_{RN}}^{\dag}\} \rightarrow \{{{\mathcal{Q}}_{RN'}}^{\dag},{{\mathcal{Q}}_{RN'}}^{\dag}\} \\ \notag &
\{{{\mathcal{Q}}_{RN}}^{\dag},{{\mathcal{Q}}_{RN}}^{\dag}\}\rightarrow \{{{\mathcal{Q}}_{RN'}}^{\dag},{{\mathcal{Q}}_{RN'}}^{\dag}\},{\,}{\,}{\,}\{{{\mathcal{Q}}_{RN'}}^{\dag},{{\mathcal{Q}}_{RN'}}^{\dag}\}\rightarrow \{{{\mathcal{Q}}_{RN}}^{\dag},{{\mathcal{Q}}_{RN}}^{\dag}\}  \\ \notag &
\{{{\mathcal{Q}}_{RN}}^{\dag},{{\mathcal{Q}}_{RN'}}^{\dag}\}\rightarrow \{{{\mathcal{Q}}_{RN'}}^{\dag},{{\mathcal{Q}}_{RN}}^{\dag}\},{\,}{\,}{\,} \{{{\mathcal{Q}}_{RN'}}^{\dag},{{\mathcal{Q}}_{RN}}^{\dag}\}\rightarrow \{{{\mathcal{Q}}_{RN}}^{\dag},{{\mathcal{Q}}_{RN'}}^{\dag}\} \\ \notag &
\{{{\mathcal{Q}}_{RN}},{{\mathcal{Q}}_{RN'}}\} \rightarrow \{{{\mathcal{Q}}_{RN'}},{{\mathcal{Q}}_{RN}}\},{\,}{\,}{\,}\{{{\mathcal{Q}}_{RN'}},{{\mathcal{Q}}_{RN}}\} \rightarrow \{{{\mathcal{Q}}_{RN}},{{\mathcal{Q}}_{RN'}}\} \\ \notag &
[{{\mathcal{Q}}_{RN'}},{{\mathcal{Q}}_{RN}}] \rightarrow [{{\mathcal{Q}}_{RN}},{{\mathcal{Q}}_{RN'}}],{\,}{\,}{\,}[{{\mathcal{Q}}_{RN}},{{\mathcal{Q}}_{RN'}}] \rightarrow [{{\mathcal{Q}}_{RN'}},{{\mathcal{Q}}_{RN}}] \\ \notag &
[{{\mathcal{Q}}_{RN}}^{\dag},{{\mathcal{Q}}_{RN'}}^{\dag}]\rightarrow [{{\mathcal{Q}}_{RN'}}^{\dag},{{\mathcal{Q}}_{RN}}^{\dag}],{\,}{\,}{\,}[{{\mathcal{Q}}_{RN'}}^{\dag},{{\mathcal{Q}}_{RN}}^{\dag}]\rightarrow [{{\mathcal{Q}}_{RN}}^{\dag},{{\mathcal{Q}}_{RN'}}^{\dag}]  \\ \notag &
[{{\mathcal{Q}}_{RN'}},{{\mathcal{Q}}_{RN'}}] \rightarrow [{{\mathcal{Q}}_{RN}},{{\mathcal{Q}}_{RN}}],{\,}{\,}{\,}[{{\mathcal{Q}}_{RN}},{{\mathcal{Q}}_{RN}}] \rightarrow [{{\mathcal{Q}}_{RN'}},{{\mathcal{Q}}_{RN'}}] \\ \notag &
[{{\mathcal{Q}}_{RN'}}^{\dag},{{\mathcal{Q}}_{RN'}}^{\dag}] \rightarrow [{{\mathcal{Q}}_{RN}}^{\dag},{{\mathcal{Q}}_{RN}}^{\dag}],{\,}{\,}{\,}[{{\mathcal{Q}}_{RN}}^{\dag},{{\mathcal{Q}}_{RN}}^{\dag}] \rightarrow [{{\mathcal{Q}}_{RN'}}^{\dag},{{\mathcal{Q}}_{RN'}}^{\dag}]
\end{align}
Consequently, the $N=4$, $d=1$ algebra of relation (\ref{commutatorsanticomm}) remains invariant under the transformation (\ref{transfduality}). Note that the solutions of the fermionic equations of motion, namely $\psi_1,\psi_2$ originate from spinors that can be written as \cite{oikonomoureissner} $\Psi_i=\psi_ie^{-i\omega t+k_xx}$ with $i=1,2$. Therefore the transformation (\ref{transfduality}) corresponds to a $\mathcal{C}\mathcal{P}\mathcal{T}$ transformation of the initial fermionic system at hand.

\subsection{A Brief Comment on Topological Charges and Supersymmetry}
  
The topological charges often occur in supersymmetric algebras, as was first noticed in reference \cite{wittentplc}, in which reference the  topological charge terminology was first used. In addition, in reference \cite{fayet} it was noticed that the topological charges cannot be considered as central charges, since these operators do not actually commute with all the operators and supercharges that constitute the superalgebra. Furthermore, as was also noted in \cite{fayet}, the topological charges are actually symmetries of the quantum field theory but not symmetries (Noether symmetries) of the $S$-matrix of the theory. The supersymmetries with topological charges we found in this article seem to fall in that category, that is, these are symmetries of the quantum theory and not of the $S$-matrix. These topological charges we found, presumably indicate some additional external symmetry of the quantum field theory that describes the fermionic modes in the curved Reissner-Nordstr\"{o}m-anti-de Sitter background. 

We have to note that the theoretical framework used in \cite{wittentplc} was actually a supersymmetric algebra in the presence of topological defect. As was evinced \cite{wittentplc}, the presence of topological charges in such frameworks are unavoidable, however in our case the existence of topological charges in our case is probably due to the specific curved black hole background. Moreover, as was pointed out in reference \cite{topologicalcharges1}, finding a non ''central'' charge that does not commute with the rest operators of the superalgebra, indicates the existence of a richer, possibly non-linear, supersymmetric underlying structure. It would be interesting enough to find whether such a structure exists in the case of our fermionic system in curved Reissner-Nordstr\"{o}m-anti-de Sitter and other curved backgrounds, but we defer this rather involved issue to a future work. Let us note finally that non-commuting topological charges are not a strange feature in supersymmetric theories, since these often occur in string theory contexts. For example, as we already mentioned, in a Ad$S_5\times S^5$ D-brane background with superalgebra $su(2,2|4)$ \cite{topologicalcharges2} there exist many topological charges with this feature, that is, having non-trivial commutation relations with the elements of the superalgebra.

\section{Localized Fermions and SUSY QM-Trivial Central Charge Case}

In this section we shall present a way of generalizing each $N=2$, $d=1$ algebra of the fermionic system to have a trivial real central charge. Following \cite{haag}, we consider the central charge to be an operator equal to the anticommutator of two Fermi charges and in addition that it commutes with every element of the superalgebra. Let us focus on one of the two $N=2$, $d=1$ superalgebras, namely the one described by the complex supercharge $\mathcal{Q}_{RN }$. The algebra of relation (\ref{structureqns}) can be easily extended to include a real central charge, denoted by $\mathcal{Z}$, in the following way:
\begin{equation}\label{s7supcghagemixcentch}
\mathcal{Q}_{Z}=\bigg{(}\begin{array}{ccc}
  -\eta & 0 \\
  \mathcal{D}_{RN } & \eta  \\
\end{array}\bigg{)},{\,}{\,}{\,}\mathcal{Q}_{Z}=\bigg{(}\begin{array}{ccc}
  -\eta &  \mathcal{D}_{RN }^{\dag} \\
 0 & \eta \\
\end{array}\bigg{)}
\end{equation}
with ''$\eta$'' being an arbitrary $2\times 2$ real matrix. The Hamiltonian of the supersymmetric quantum mechanical system in this case reads,
\begin{equation}\label{s11fgghhfcentcharge}
\mathcal{H}_{Z}=\bigg{(}\begin{array}{ccc}
 \mathcal{D}_{RN}\mathcal{D}_{RN }^{\dag}+2\eta^2 & 0 \\
  0 & \mathcal{D}_{RN }^{\dag}\mathcal{D}_{RN }+2\eta^2  \\
\end{array}\bigg{)}
\end{equation}
The real central charge $\mathcal{Z}$ of the centrally extended $N=2$, $d=1$ SUSY QM algebra is equal to the following operator,
\begin{equation}\label{centrextendcc}
\mathcal{Z}=\bigg{(}\begin{array}{ccc}
  2\eta^2 & 0 \\
  0 & 2\eta^2  \\
\end{array}\bigg{)}.
\end{equation}
Consequently, the new form of the extended $N=2$ SUSY QM with trivial central charge algebra is described by the following commutation-anticommutation relations:
\begin{equation}\label{relationsforsusycc}
\{\mathcal{Q}_{Z},\mathcal{Q}^{\dag}_{Z}\}=\mathcal{H}_{Z}{\,}{\,},\{\mathcal{Q}_{Z},\mathcal{Q}_{Z}\}=\mathcal{Z},{\,}{\,}\{\mathcal{Q}^{\dag}_{Z},\mathcal{Q}^{\dag}_{Z}\}=\mathcal{Z},{\,}{\,}{\,}[\mathcal{H}_{Z},\mathcal{Q}_{Z}]=[\mathcal{H}_{Z},\mathcal{Q}_{Z}^{\dag}]=0
\end{equation}
One question that springs to mind is how this central charge modifies the properties of the supersymmetric quantum mechanical system. Considering Hilbert space states, the new supercharges no longer map the parity-even to parity-odd states, when non-zero modes of the system are taken into account. More importantly, when the matrix $2\eta^{2}$ is an odd compact matrix, the index of the operator $\mathcal{D}_{RN }$ is invariant under the compact perturbations generated by the matrix $2\eta^2$. In order to see this we shall use a theorem for trace-class operators. Let $\mathcal{C}$ an odd compact symmetric matrix \cite{thaller} of the form:
\begin{equation}\label{susyqmrnmassive}
\mathcal{C}=\left(%
\begin{array}{cc}
 0 & \mathcal{C}_1
 \\ \mathcal{C}_2 & 0\\
\end{array}%
\right)
\end{equation}
Consider that the operator $\mathcal{D}$ is a trace-class operator and also an operator $\mathcal{D}_{p}$ which is equal to $\mathcal{D}_{p}=\mathcal{D}+\mathcal{C}$. Then $\mathcal{D}_{p}$ is a compact perturbation of the operator $\mathcal{D}$ and since the operator $\mathrm{tr}\mathcal{W}e^{-t(\mathcal{D}+\mathcal{C})^2}$ is trace class (owing to the fact that compact perturbations of trace-class operators are also trace-class operators), the following theorem holds true (see \cite{thaller} page 168, Theorem 5.28),
\begin{equation}\label{indperturbatrn1}
\mathrm{ind}\mathcal{D}_p=\mathrm{ind}(\mathcal{D}+\mathcal{C})=\mathrm{ind}\mathcal{D}
,\end{equation}
with $\mathcal{C}$ the symmetric odd operator of the form (\ref{susyqmrnmassive}). Coming back to our case, when the matrix $\eta^2$ is compact and odd, theorem (\ref{indperturbatrn1}) holds true and therefore, the following relations hold true: 
\begin{align}\label{genkerrelationscompnew}
&-\Delta = \mathrm{dim}{\,}\mathrm{ker}\mathcal{D}_{RN }^{\dag}\mathcal{D}_{RN }-\mathrm{dim}{\,}\mathrm{ker}\mathcal{D}_{RN }\mathcal{D}_{RN}^{\dag}=
\\ \notag & \mathrm{dim}{\,}\mathrm{ker}(\mathcal{D}_{RN}^{\dag}\mathcal{D}_{RN }+2\eta^2)-\mathrm{dim}{\,}\mathrm{ker}(\mathcal{D}_{RN}\mathcal{D}_{RN}^{\dag}+2\eta^2)
\end{align}
with $\Delta$ the Witten index of the $N=2$ supersymmetric system without central charge. An immediate consequence of relation (\ref{genkerrelationscompnew}) is that the Witten index $\Delta$ of the non-central charge supersymmetric quantum mechanical system (\ref{structureqns}) is invariant under the perturbations caused by the central charge extension of the system. This result is valid when real central charges are taken into account.

Before closing, let us see how the constraints ''compact'' and ''odd'' modify the matrix $2\eta^2$. Compact means that the matrix $\eta$ must contain elements which are finite numbers or fast decaying functions and odd means that $\eta^2$ must have the following form:
\begin{equation}\label{etasquareoddcccc}
\eta^2=\bigg{(}\begin{array}{ccc}
  0 & b \\
  -b & 0  \\
\end{array}\bigg{)},
\end{equation}
In order this to be true, the matrix $\eta$ must be of the form:
\begin{equation}\label{etasquareoddcccceta}
\eta=\bigg{(}\begin{array}{ccc}
  \sqrt{b} & -\sqrt{b} \\
  \sqrt{b} & \sqrt{b}  \\
\end{array}\bigg{)}.
\end{equation}
with $b$ some positive integer.

Of course we have to notice that the extra structure presented in this section is an extra artificial structure having nothing to do with the supersymmetric structure presented in the previous section. We just showed that we can have trivial topological charges in the initial $N=2$ algebra of section 1.

\section*{Concluding Remarks}

In this paper we demonstrated that the fermions in a Reissner-Nordstr\"{o}m-anti-de Sitter black hole background apart from two $N=2$, $d=1$ supersymmetries, also possess an $N=4$, $d=1$ supersymmetry with non-trivial topological charges. Although the initial fermionic system had no inherent global supersymmetry, the field theory has a rich non-trivial supersymmetric underlying structure, that can get even more involved when the number of fermion flavors increases. This behavior has been pointed out previously in the literature, see for example \cite{oiko1}. It seems that actually global supersymmetry plays no obvious role in the issues we studied in this article and the supersymmetry we found is an underlying symmetry of the field theory, a symmetry that is not a symmetry of the S-matrix of the theory. More generally, spacetime supersymmetry in $d>1$ dimensions and SUSY QM, which is an one dimensional supersymmetry algebra, are in principle different concepts, with the only 
 possible connection being the fact that extended (with $N = 4, 6...$) SUSY QM models can be obtained by $N = 2$ and $N = 1$ Super-Yang Mills models, by dimensionally reducing them down to one dimension \cite{ivanov1,ivanov2}. However, the complex supercharges of $N = 2$, $d=1$ and also of $N=4$, $d=1$ SUSY QM are not related to the generators of spacetime supersymmetry and thereby, SUSY QM does not directly relate fermions and bosons, with fermions and bosons considered as representations of the super-Poincare graded Lie algebra in four dimensions. What the SUSY QM supercharges actually do is that they render the Hilbert space of quantum states a $Z_2$-graded vector space, in the simplest case of $N=2$ SUSY QM. Moreover, these supercharges generate the transformations between the Witten parity eigenstates. As was already noted in the present paper, the $N=4$ one dimensional supersymmetry is not a spacetime symmetry nor a symmetry of the S-matrix, but a symmetry of the field theory in
  terms of the equations of motion. A symmetry that reveals a possible even more involved symmetry yet to be found, due to the existence of non-trivial topological charges that do not commute with the rest of the operators of the SUSY QM algebra.

\end{document}